\documentclass[12pt,thmsa]{article}
\usepackage{amssymb}

\usepackage{sw20lart}

\input{tcilatex}
\begin{document}

\author{Emilio Santos \and Departamento de F\'{i}sica. Universidad de Cantabria.
Santander. Spain}
\title{Neutron stars in generalized f(R) gravity}
\date{February, 7, 2012 }
\maketitle

\begin{abstract}
Quartic gravity theory is considered with the Einstein-Hilbert Lagrangean $%
R+aR^{2}+bR_{\mu \nu }R^{\mu \nu },$ $R_{\mu \nu }$ being Ricci\'{}s tensor
and R the curvature scalar. The parameters $a$ and $b$ are taken of order 1
km$^{2}.$ Arguments are give which suggest that the effective theory so
obtained may be a plausible approximation of a viable theory. A numerical
integration is performed of the field equations for a free neutron gas. The
result is that the star mass increases with increasing central density until
about 1 solar mass and then decreases. The baryon number increases
monotonically, which suggests that the theory allows stars in equilibrium
with arbitrary baryon number, no matter how large.

\end{abstract}

\section{ Introduction}

General relativity has passed all observation tests so far, but the real
theory of gravity may well differ significantly from it in strong field
regions. In fact conceptual difficulties in quantizing Einstein's theory and
astrophysical observations suggest that general relativity may require
modifications. In recent years a great effort has been devoted to the study
of extended gravity theories, mainly with the goal of finding physical
explanations to the accelerated expansion of the universe and other
astrophysical observations, like the flat rotation curves in galaxies \cite
{Odintsov}, \cite{Cap}.

Compact stars are an ideal natural laboratory to look for possible
modifications of Einsteins theory and their observational signatures. A
rather general class of alternative theories of gravity has been considered
recently\cite{Pani} to study slowly rotating compact stars with the purpose
of investigating constraints on alternative theories. Several studies of
compact stars, in particular neutron stars, have been made within extended
gravity theories\cite{Cap2}, \cite{Maeda}, \cite{Babichev}, \cite{Laurentis}%
. There are also theories which prevent the appareance of singularities like
``gravastars''\cite{Mazur} and Eddington inspired gravity\cite{Pani1}.

The most popular modification of general relativity, since the early days of
general relativity\cite{Schmidt}, derives from an extension of the
Einstein-Hilbert action of the form 
\begin{equation}
S=\frac{1}{2k}\int d^{4}x\sqrt{-g}(R+F)+S_{mat},  \label{n1}
\end{equation}
where $k$ is $8\pi $ times Newton\'{}s constant and I use units $c=1$
throughout, and $F$ is a function of the scalars which may be obtained by
combining the Riemann tensor, $R_{\mu \nu \lambda \sigma },$ and its
derivatives, with the metric tensor, $g_{\mu \nu }$. In particular the
theory derived from the choice $F(R),$ where $R$ is the Ricci scalar, has
been extensively explored under the name of \textit{f(R)-gravity}\cite
{Faraoni},\cite{de Felice}. More general is fourth order gravity\cite{Cap1},
which derives from the choice 
\begin{equation}
F=F\left( R,R_{\mu \nu }R^{\mu \nu },R_{\mu \nu \lambda \sigma }R^{\mu \nu
\lambda \sigma }\right) ,  \label{1a}
\end{equation}
$R_{\mu \nu }$ being the Ricci tensor. A particular example of eq.$\left( 
\ref{1a}\right) $ is the quadratic Lagrangian which may be written without
loss of generality 
\begin{equation}
F=aR^{2}+bR_{\mu \nu }R^{\mu \nu }.  \label{1b}
\end{equation}
(Riemann square does not appear because it may be eliminated using the the
Gauss-Bonnett combination 
\[
R_{GB}^{2}\equiv R^{2}-4R_{\mu \nu }R^{\mu \nu }+R_{\mu \nu \lambda \sigma
}R^{\mu \nu \lambda \sigma }, 
\]
which does not contribute to the field equations in a quadratic Lagrangian.)
The Newtonian limit of the field equations derived from the Lagrangian eq.$%
\left( \ref{1b}\right) $ has been studied elsewhere\cite{Santos}, \cite
{Stabile}.

In this paper I report on a calculation of neutron stars using the theory
derived from eq.$\left( \ref{1b}\right) $ with the particular choice of
parameters $b=-2a,$ $\sqrt{a}=1$ km. That theory is apparently not viable
for two reasons. Firstly, in order not contradicting Solar System and
terrestrial observations the parameters $a$ and $b$ should be not greater
than a few millimeters. Secondly the weak field limit of the theory should
not present ghosts\cite{Hindawi}, \cite{de Felice2}. A solution to both
problems is to assume that eq.$\left( \ref{1b}\right) $ is an approximation,
valid for the strong fields appearing in neutron stars, of another function $%
F$ which is extremely small in the weak field limit. This would be the case,
for instance, if $F$ has the form

\begin{equation}
F=aR^{2}+bR_{\mu \nu }R^{\mu \nu }-c\log \left( 1+(a/c)R^{2}+(b/c)R_{\mu \nu
}R^{\mu \nu }\right) .  \label{9}
\end{equation}
with $a\simeq 10^{6}m^{2},b-2a,c=1/(10^{26}m^{2}).$ Thus eq.$\left( \ref{9}%
\right) $ may be approximated by 
\begin{equation}
F\simeq \frac{1}{2c}\left( aR^{2}+bR_{\mu \nu }R^{\mu \nu }\right) ^{2}\text{
}\lesssim 10^{-32}R\text{ },  \label{10}
\end{equation}
for the Solar System and the relative error due to the terms neglected in
going from eq.$\left( \ref{9}\right) $ to eq.$\left( \ref{10}\right) $ is
smaller than $10^{-12}$. I have taken into account that $R^{2}\sim R_{\mu
\nu }R^{\mu \nu }\sim (k\rho )^{2}$ and that the typical density $\rho \sim
10^{4}$ kg/m$^{3}$. The inequality in $\left( \ref{10}\right) $ shows that
the correction to GR due to the function $F$, eq.$\left( \ref{9}\right) ,$
is neglible in Solar System or terrestrial calculations. Also the problem of
the ghosts in the weak field limit disappears with that choice. Indeed the
theory is fine in the context of low-energy effective actions because the
contribution of $R_{\mu \nu }R^{\mu \nu }$ is so small that it never
dominates the dynamics of the background. On the other hand the $R^{2}$ term
does not introduce extra graviton modes.

In contrast in neutron stars, where $\rho \sim 10^{18}$ kg/m$^{3},$ the
latter (logaritmic) term of eq.$\left( \ref{9}\right) $ is about $10^{-12}$
times the former terms and it may be ignored in the calculation.

\section{Field equations}

The tensor field equation derived from the functional eqs.$\left( \ref{n1}%
\right) $ and the latter eq$.\left( \ref{10}\right) $ may be taken from the
literature\cite{Odintsov1}, \cite{Turner}. I shall write it in terms of the
Einstein tensor, $G_{\mu \nu }$, rather than the Ricci tensor, $R_{\mu \nu
}, $ and in a form that looks like the standard Einstein equation of general
relativity eq.$\left( \ref{E}\right) $. That is 
\begin{equation}
G_{\mu \nu }\equiv R_{\mu \nu }-\frac{1}{2}g_{\mu \nu }R=k\left( T_{\mu \nu
}^{mat}+T_{\mu \nu }^{ef}\right) ,  \label{E}
\end{equation}

\begin{eqnarray}
&&kT_{\mu \nu }^{ef}\equiv -(2a+b)\left[ \nabla _{\mu }\nabla _{\nu
}G-g_{\mu \nu }\Box G\right] -2\left( a+b\right) \left[ -GG_{\mu \nu }+\frac{%
1}{4}g_{\mu \nu }G^{2}\right]  \nonumber \\
&&-b\left[ 2G_{\mu }^{\sigma }G_{\sigma \nu }-\frac{1}{2}g_{\mu \nu
}G_{\lambda \sigma }G^{\lambda \sigma }-\nabla _{\sigma }\nabla _{\nu
}G_{\mu }^{\sigma }-\nabla _{\sigma }\nabla _{\mu }G_{\nu }^{\sigma }+\Box
G_{\mu \nu }\right] .  \label{1}
\end{eqnarray}

We are interested in static problems of spherical symmetry and will use the
standard metric

\begin{equation}
ds^{2}=-\exp \left( \beta \left( r\right) \right) dt^{2}+\exp \left( \alpha
\left( r\right) \right) dr^{2}+r^{2}d\Omega ^{2}.  \label{metric}
\end{equation}
Thus $G_{\mu \nu }\left( r\right) $ and $T_{\mu \nu }^{mat}\left( r\right) $
have 3 independent components each, so that including $\alpha \left(
r\right) $ and $\beta \left( r\right) $ there are 8 variables. On the other
hand there are 8 equations, namely 3 eqs$\left( \ref{1}\right) ,$ 3 more
equations giving the independent components of $G_{\mu \nu }$ in terms of $%
\alpha $ and $\beta $ and 2 equations of state relating the 3 independent
components of $T_{\mu \nu }^{mat}.$ I shall assume local isotropy for matter
so that one of the latter will be the equality $T_{11}^{mat}=T_{22}^{mat}$ ($%
=T_{33}^{mat}$ in spherical symmetry.) In principle the remaining 7 coupled
non-linear equations may be solved exactly by numerical methods, as will be
explained in Section 4.

Before proceeding, a note about the signs convention is in order. As is well
known different authors use different signs in the definition of the
relevant quantities. Here I shall make a choice which essentially agrees
with the one of Ref.\cite{Faraoni}. It may be summarized as follows

\begin{equation}
g_{00}=-\exp \beta ,G_{\mu \nu }=R_{\mu \nu }-\frac{1}{2}g_{\mu \nu
}R=kT_{\mu \nu },T_{0}^{0}=-\rho .  \label{signs}
\end{equation}
After that I shall write the three independent components of eq.$\left( \ref
{1}\right) $using the notation 
\begin{eqnarray}
T_{0}^{0} &=&-\rho ,T_{1}^{1}=p,T_{2}^{2}=q,T_{\mu }^{\mu }=T=p+2q-\rho , 
\nonumber \\
\left( T_{mat}\right) _{0}^{0} &=&-\rho _{mat},\left( T_{mat}\right)
_{1}^{1}=\left( T_{mat}\right) _{2}^{2}=\left( T_{mat}\right)
_{3}^{3}=p_{mat}.  \label{ropq}
\end{eqnarray}
In the following I will name $\rho ,p$ and $q$ the \textit{total} density,
radial pressure and transverse pressure respectively, whilst $\rho _{mat}$
and $p_{mat}$ will be named \textit{matter} density and pressure
respectively (remember that we assume local isotropy for matter, that is the
equality of radial and transverse matter pressures.) The differences $\rho
-\rho _{mat},p-p_{mat}$ and $q-p_{mat}$ will be named \textit{effective}
density, radial pressure and transverse pressure respectively.

After some algebra I get for the components of the tensor eq.$\left( \ref{1}%
\right) $

\begin{eqnarray}
-\rho _{mat} &=&-\rho +(2a+b)e^{-\alpha }\left[ -\frac{d^{2}T}{dr^{2}}%
-\left( \frac{2}{r}-\frac{1}{2}\alpha ^{\prime }\right) \frac{dT}{dr}\right]
+(a+b)k(\frac{1}{2}T^{2}+2T\rho )  \nonumber \\
&&+b\exp (-\alpha )\left[ -\frac{2\beta ^{\prime }}{r}\left( q-p\right)
+\left( \frac{1}{2}\alpha ^{\prime }\beta ^{\prime }-\beta ^{\prime \prime }-%
\frac{2\beta ^{\prime }}{r}\right) \left( \rho +p\right) \right]  \nonumber
\\
&&+b\left[ -\Delta \rho +2k\rho ^{2}-\frac{1}{2}k\left[ \rho
^{2}+p^{2}+2q^{2}\right] \right] ,\smallskip  \label{rom}
\end{eqnarray}
\begin{eqnarray}
p_{mat} &=&p-(2a+b)e^{-\alpha }\left( \frac{2}{r}+\frac{1}{2}\beta ^{\prime
}\right) \frac{dT}{dr}+(a+b)k(\frac{1}{2}T^{2}-2Tp)  \nonumber \\
&&+b\left[ \Delta p+2kp^{2}-\frac{1}{2}k\left[ \rho ^{2}+p^{2}+2q^{2}\right]
\right]  \nonumber \\
&&+b\exp (-\alpha )\left[ \left( \frac{2\alpha ^{\prime }}{r}+\frac{4}{r^{2}}%
\right) \left( q-p\right) +\left( -\frac{1}{2}\alpha ^{\prime }\beta
^{\prime }+\beta ^{\prime \prime }\right) (\rho +p)\right] ,  \label{pm}
\end{eqnarray}
\begin{eqnarray}
p_{mat} &=&q-(2a+b)e^{-\alpha }\left[ \frac{d^{2}T}{dr^{2}}+\left( \frac{1}{r%
}+\frac{1}{2}\beta ^{\prime }-\frac{1}{2}\alpha ^{\prime }\right) \frac{dT}{%
dr}\right]  \nonumber \\
&&+(a+b)k(\frac{1}{2}T^{2}-2Tq)+b\left[ \Delta q+2kq^{2}-\frac{1}{2}k\left[
\rho ^{2}+p^{2}+2q^{2}\right] \right]  \nonumber \\
&&+b\exp (-\alpha )\left[ \left( -\frac{\alpha ^{\prime }}{r}+\frac{\beta
^{\prime }}{r}-\frac{2}{r^{2}}\right) \left( q-p\right) +\frac{\beta
^{\prime }}{r}(\rho +p)\right] .  \label{qm}
\end{eqnarray}
Addition of these 3 equations gives the trace equation, that is 
\begin{equation}
T_{mat}\equiv 3p_{mat}-\rho _{mat}=T-\left( 6a+2b\right) \Delta T,
\label{trace}
\end{equation}
where $\Delta $ is the Laplacean operator in curved space-time 
\begin{equation}
\Delta \equiv \exp (-\alpha )\left[ \frac{d^{2}}{dr^{2}}+\left( \frac{2}{r}+%
\frac{1}{2}\beta ^{\prime }-\frac{1}{2}\alpha ^{\prime }\right) \frac{d}{dr}%
\right] .  \label{laplace}
\end{equation}
The quantities G$_{\mu }^{\nu }$ are related to the metric coefficients $%
\alpha $ and $\beta $ and their derivatives (see e.g. \cite{Synge}), hence
to $\rho ,p$ and $q,$ that is 
\begin{eqnarray}
\exp (-\alpha ) &=&1-\frac{2m}{r},\frac{\alpha ^{\prime }}{2}=\frac{m-4\pi
\rho r^{3}}{r^{2}-2mr},\beta ^{\prime }=2\frac{m+4\pi r^{3}p}{r^{2}-2mr}, 
\nonumber \\
\beta ^{\prime \prime } &=&\frac{8\pi r^{2}\left( r\rho +rp+3p^{\prime
}\right) }{r^{2}-2mr}-\frac{4\left( m+4\pi r^{3}p\right) \left( r-m-4\pi
r^{3}\rho \right) }{\left( r^{2}-2mr\right) ^{2}},  \label{alfabeta}
\end{eqnarray}
where I have used units $k=8\pi ,c=1$ and the radial derivative of $\alpha
\left( \beta ^{\prime }\right) $ is labelled $\alpha ^{\prime }\left( \beta
^{\prime \prime }\right) $. The mass parameter $m$ is defined by 
\begin{equation}
m=\int_{0}^{r}4\pi x^{2}\rho (x)dx.  \label{mass}
\end{equation}
The condition that Einstein tensor, $G_{\mu \nu },$ is divergence free leads
to the hydrostatic equilibrium equation, that is 
\begin{equation}
\frac{dp}{dr}=\frac{2(q-p)}{r}-\frac{1}{2}\beta ^{\prime }\left( \rho
+p\right) .  \label{OV}
\end{equation}

\section{Application to neutron stars}

For neutron stars, when are quadratic gravity corrections relevant?. In
order to answer this question we should estimate the conditions where $%
T_{\mu \nu }^{ef},$ eq.$\left( \ref{1}\right) ,$ is comparable to$T_{\mu \nu
}^{mat}.$ Terms with derivatives are of order 
\[
a\Box G\sim (a/R_{0}^{2})G, 
\]
$R_{0}$ being the radius of the hypothetical star. Thus these terms are
relevant if the dimensionless quantity $a/R_{0}^{2}$ is of order unity,
which implies that $a$ and $b$ should be of order the star radius, that is a
few kilometers. Terms without derivatives are of order 
\[
aG^{2}\sim (ak\rho /c^{2})G, 
\]
similar to those with derivatives$.$

In order to solve eqs.$\left( \ref{rom}\right) $ to $\left( \ref{OV}\right) $
we need an equation of state (\textit{eos}), that is a relation between $%
p_{mat}$ and $\rho _{mat}$, appropriate for a system of neutrons. For the
calculation here reported I shall choose the eos of a free (non-interacting)
neutron gas. In order to make easier the rather involved numerical
integration of the equations, I will simplify the said eos writing 
\begin{equation}
\rho _{mat}=3p_{mat}+Cp_{mat}^{3/5},\text{ }C=2.34,  \label{eos}
\end{equation}
where $\rho _{mat}$ and $p_{mat}$ are in units of $7.2\times 10^{18}$ kg m$%
^{-3}$. This equation is correct in the limit of high density, where $\rho
_{mat}\simeq 3p_{mat}$, and has the same dependence $p_{mat}\propto \rho
_{mat}^{5/3}$ as the eos of the free neutron gas in the nonrelativistic
limit of low density. The constant $C$ is so chosen that we get the same
result as Oppenheimer and Volkoff\cite{OV} \ for the maximum mass stable
star in their general relativistic calculation.

A relevant quantity is the baryon number of the star, $N$, which may be
calculated from the baryon number density $n(r)$ via 
\begin{equation}
N=\int_{0}^{R}\frac{n(r)}{\sqrt{1-2m(r)/r}}4\pi r^{2}dr,  \label{N}
\end{equation}
in our units. A relation between the number density and the matter density
(or pressure) may be got from the solution of the equation 
\[
p_{mat}=n\frac{d\rho _{mat}}{dn}-\rho _{mat}, 
\]
which follows from the first law of thermodynamics. Inserting eq.$\left( \ref
{eos}\right) $ here we get a differential equation which may be easily
solved with the condition $\rho _{mat}/n\rightarrow \mu $ for $\rho
\rightarrow 0$, $\mu $ being the neutron mass. I get 
\begin{equation}
n=C^{5/8}p_{mat}^{3/5}\left( 4p_{mat}^{2/5}+C\right) ^{3/8},  \label{n}
\end{equation}
where the unit of baryon number density is $\mu ^{-1}7.2\times 10^{18}$ kg m$%
^{-3}$.

\section{Neutron stars in extended gravity}

In order to derive the structure of neutron stars in generalized f(R)
gravity theory, as defined by eqs.$\left( \ref{1}\right) ,$ we should solve
the coupled eqs.$\left( \ref{rom}\right) $ to $\left( \ref{N}\right) $ plus
the hydrostatic equilibium eq.$\left( \ref{OV}\right) $ with our choice of
the parameters $a$ and $b$, namely $b=-2a,$ $\sqrt{a}=0.96$ km. This choose
of $a$ and $b$ makes the calculation specially simple.

We need just 3 amongst the 4 eqs.$\left( \ref{rom}\right) $ to $\left( \ref
{trace}\right) ,$ because only 3 are independent. I choose eqs.$\left( \ref
{trace}\right) ,$the difference eq.$\left( \ref{qm}\right) $ minus eq.$%
\left( \ref{pm}\right) ,$ and eq.$\left( \ref{pm}\right) $, which may be
rewritten 
\begin{equation}
\frac{dT}{dr}=T^{\prime },\frac{dT^{\prime }}{dr}=-\left( \frac{2}{r}+\frac{1%
}{2}\beta ^{\prime }-\frac{1}{2}\alpha ^{\prime }\right) \frac{dT}{dr}+\frac{%
T-T_{mat}}{2a},  \label{dT}
\end{equation}

\begin{eqnarray}
\frac{dh}{dr} &=&h^{\prime },h\equiv q-p,  \nonumber \\
\frac{dh^{\prime }}{dr} &=&-\left( \frac{2}{r}+\frac{1}{2}\beta ^{\prime }-%
\frac{1}{2}\alpha ^{\prime }\right) h^{\prime }+\exp \alpha \left[ \frac{h}{%
2a}+kTh-2k(h+2p)h\right]  \nonumber \\
&&-\left[ \left( -\frac{3\alpha ^{\prime }}{r}+\frac{\beta ^{\prime }}{r}-%
\frac{6}{r^{2}}\right) h+(\frac{\beta ^{\prime }}{r}-\beta ^{\prime \prime }+%
\frac{1}{2}\alpha ^{\prime }\beta ^{\prime })(\rho +p)\right] ,  \label{dh}
\end{eqnarray}

\begin{eqnarray}
p_{mat} &=&p+ak(2Tp-\frac{1}{2}T^{2}-3p^{2}+\rho ^{2}+2q^{2})  \nonumber \\
&&-2a\exp (-\alpha )\left[ \Delta p+\left( \frac{2\alpha ^{\prime }}{r}+%
\frac{4}{r^{2}}\right) h+\left( \beta ^{\prime \prime }-\frac{1}{2}\alpha
^{\prime }\beta ^{\prime }\right) (\rho +p)\right] ,  \label{dp}
\end{eqnarray}
where the Laplacean operator $\Delta $ was defined in eq.$\left( \ref
{laplace}\right) .$ Finally we need the hydrostatic equilibrium eq.$\left( 
\ref{OV}\right) $.

The numerical calculation goes as follows. From the values of all variables
at a given radial coordinate $r$, integration of the linear differential eqs.%
$\left( \ref{dT}\right) ,$ $\left( \ref{dh}\right) $ and $\left( \ref{OV}%
\right) ,$ taking eq.$\left( \ref{mass}\right) $ into account, provides the
values of $m,T,T^{\prime },h,h^{\prime }$ and $p$ at $r+dr.$ Hence the
relation (see eq.$\left( \ref{ropq}\right) )$%
\[
\rho =p+2q-T=2h+3p-T, 
\]
gives $\rho \left( r+dr\right) ,$ whence eq.$\left( \ref{OV}\right) $ gives $%
p^{\prime }\left( r+dr\right) $ which allows obtaining $\rho ^{\prime
}\left( r+dr\right) .$ After that we have all quantities needed to get $%
d^{2}p/dr^{2}$ from the derivative of eq.$\left( \ref{OV}\right) ,$ that is 
\[
\frac{d^{2}p}{dr^{2}}=\frac{2h^{\prime }}{r}-\frac{2h}{r^{2}}-\frac{1}{2}%
\beta ^{\prime \prime }\left( \rho +p\right) -\frac{1}{2}\beta ^{\prime
}\left( \rho ^{\prime }+p^{\prime }\right) . 
\]
Hence we get $p_{mat}$ from eq.$\left( \ref{dp}\right) $ taking eqs.$\left( 
\ref{laplace}\right) $ and $\left( \ref{alfabeta}\right) $ into account,
which allows obtaining $\rho _{mat}$ via the eos eq.$\left( \ref{eos}\right)
,$ whence $T_{mat}=3p_{mat}-\rho _{mat}$ follows (remember that we assume
local isotropy for matter, that is $p_{mat}=q_{mat}$.) In this way we obtain
all the quantities at $r+dr$ and the process may be repeated in order to get
the quantities at $r+2dr$ , and so on. This shows that our equations form a
consistent system.

As initial conditions for the differential equations we need the values of
the following variables at the origin: $T\left( 0\right) ,h\left( 0\right)
,p\left( 0\right) ,T^{\prime }\left( 0\right) ,h^{\prime }\left( 0\right) .$
The latter 2 should be taken equal to zero in order that there is no
singularity, and $h\left( 0\right) =0$ because there is no distinction
between radial, $p$, and transverse pressure, $q$ at the origin. We are left
with just two free parameters, namely $p\left( 0\right) $ and $T\left(
0\right) $, but there is a constraint, that is the condition that $%
T\rightarrow 0$ for $r\rightarrow \infty .$ Indeed the matter density and
pressure are positive within the star, so that $p_{mat}\left( r\right) =0$
for any $r>R,$ $R$ being the star radius (incidentally, there is some
contribution to the star mass outside the star surface due to the effective
density.) For $r>R\left( r\right) $ the quantity $T\left( r\right) $ (and
the density $\rho \left( r\right) )$ should decrease rapidly with increasing 
$r$. As a consequence only the value of $p\left( 0\right) $ may be chosen at
will, whilst the value of $T(0)$ should be so chosen as to guarantee the
rapid decrease of $T(r)$ for $r>R$. Consequently I have been lead to perform
the integration several times for each choice of $p\left( 0\right) $, with a
different value of $T\left( 0\right) $ each time, until I got a value of $%
T\left( r\right) $ sufficiently small for large enough $r$ (that is greater
then the star radius). This procedure presents the practical difficulty that
requires a fine tuning of $T(0)$ due to the fact that for large $r>R$ the
solution of eqs.$\left( \ref{dT}\right) $ is approximately of the form 
\[
T\left( 0\right) \sim \frac{A}{r}\exp \left( \frac{r}{\sqrt{2a}}\right) +%
\frac{B}{r}\exp \left( -\frac{r}{\sqrt{2a}}\right) . 
\]
Thus the parameter $A$ should be very accurately nil in order that the first
term does not surpasses the second one at large $r$. This is specially so if
the parameter $a$ is small, and this is why I have chosen to study the case
of a relatively large value of $a.$ Also in order to alleviate the problem I
have substituted a differential equation for a new variable $f$ for the eqs.$%
\left( \ref{dT}\right) $ where 
\[
T=\frac{f}{r}\exp \left( -\frac{r}{\sqrt{2a}}\right) . 
\]
Thus the condition that $f$ remains bounded for $r\rightarrow \infty $
replaces the stronger condition that $T\rightarrow 0$ and the numerical
procedure is less unstable.

In summary we obtain a one-parameter family of equilibrium stars, one for
each value of the central total pressure $p\left( 0\right) $. Table 1
reports the results of the calculation. As in the standard (GR) theory of
neutron stars\cite{OV} the radius decreases with increasing central density,
whilst the mass increases until a maximum value and then decreases.
Therefore our theory also predicts a maximum mass for equilibrium neutron
stars. However there is a dramatic difference in the behaviour of the baryon
number, which here is always increasing with increasing central density. Of
course in stars with very large central density, matter will not be in the
form of neutrons but will consists of a mixture of different particles but I
will assume that the total baryon number is well defined. Although I have
not made a rigorous proof, the results of the calculation suggest that there
may be equilibrium configurations of neutron stars for any baryon number no
matter how large. A consequence of the strong increase of the baryon number
with a decrease of the mass implies that the binding energy becomes very
large, about 90\% of the mass in the stars with the highest central density
amongst those studied here.

Table 1 also shows that both the baryon number density, $n$, and the matter
density, $\rho _{mat},$ become very large for moderately large central total
density. This implies that the effective density, $\rho _{eff}=\rho -\rho
_{mat}$ is negative in the central region of the star although becoming
positive near and beyond the surface. However neither $\rho _{eff}$ nor $%
\rho _{mat}$ have a real physical meaning, only the total density $\rho $
being meaningful, and it remains always positive. A similar thing happens
with the pressure. The surface relative red shift is higher in our theory
than in the standard (GR) theory, but the difference is not dramatic.

\textbf{Table 1}. \textit{Our calculation. Central total pressure, }$p\left(
0\right) $\textit{, central total density, }$\rho \left( 0\right) ,$\textit{%
\ and central matter density, }$\rho _{mat}(0),$\textit{\ are in units }$%
\rho _{c}\equiv 7.2\times 10^{18}$\textit{\ kg/m}$^{3}$\textit{. Central
baryon number density, }$n(0)$\textit{, in units }$\rho _{c}/\mu ,$\textit{\ 
}$\mu $\textit{\ being the neutron mass. Star radius, }$R,$\textit{\ is in
kilometers, mass, }$M,$\textit{\ in solar masses and baryon number, }$N,$%
\textit{\ in units of solar baryon number. I report also the dimensionless
fractional surface red shift, }$\Delta \lambda /\lambda =1/\sqrt{1-2M/R}-1,$%
\textit{\ and percent binding energy, }$BE=100(N-M)/N$\textit{. An
expressions like }$1.6E2$\textit{\ means }$1.6\times 10^{2}$

$
\begin{array}{llllllll}
p\left( 0\right) & 0.01 & 0.1 & 1 & 10 & 100 & 1000 & 10000 \\ 
\rho \left( 0\right) & 0.18 & 0.82 & 4.5 & 34 & 3.1E2 & 3.0E3 & 3.0E5 \\ 
\rho _{mat}\left( 0\right) & 1.6E2 & 2.5E3 & 4.3E4 & 7.8E5 & 1.6E7 & 3.2E8 & 
6.3E9 \\ 
n\left( 0\right) & 56 & 4.5E2 & 3.7E3 & 3.3E4 & 3.2E5 & 3.0E6 & 2.8E7 \\ 
R & 10.7 & 6.7 & 4.0 & 2.7 & 2.1 & 2.2 & 2.2 \\ 
M & 0.67 & 0.80 & 0.60 & 0.39 & 0.264 & 0.268 & 0.292 \\ 
N & 0.73 & 0.94 & 1.03 & 1.13 & 1.44 & 2.00 & 2.63 \\ 
BE & 8.9\% & 15\% & 41\% & 65\% & 82\% & 87\% & 89\% \\ 
\Delta \lambda /\lambda & 0.106 & 0.22 & 0.34 & 0.31 & 0.23 & 0.23 & 0.26
\end{array}
$

\section{Discussion}

The calculations of this paper show that, if there are corrections to
general relativiy of the form of eqs.$\left( \ref{n1}\right) $ and$\left( 
\ref{1b}\right) $, then the structure of neutron stars would be dramatically
different from the one predicted by the general relativity. In particular a
new scenario would emerge for the evolution of the central region of massive
white dwarfs stars after the supernova explosion. Indeed the said central
region might contract strongly by emitting an amount of energy corresponding
to a very large fraction of the total mass. The final result will be a
neutron star with a mass maybe surpassing the believed (Oppenheimer-Volkoff)
limit. It is not possible to know how large is the new mass limit until
calculations with more realistic equations of state are performed. In
addition the predictions of the theory here considered may be quite
different for other choices of the parameters $a$ and $b$.

\section{Appendix. Neutron stars in general relativity}

For the sake of comparison with the results of our calculation using eqs.$%
\left( \ref{rom}\right) $ to $\left( \ref{n}\right) ,$ I summarize in Table
2 the results of a calculation similar to the one performed by the
Oppenheimer and Volkoff calculation\cite{OV} using general relativity. It
corresponds to taking $a=b=0$ in eqs.$\left( \ref{rom}\right) $ to $\left( 
\ref{trace}\right) $, so that $\rho =\rho _{mat},p=p_{mat}$, and I use the
eos eq.$\left( \ref{eos}\right) .$ I have extended the calculation to quite
high values of the central pressure because for those values the corrections
to $GR$ in our generalized $f(R)$ gravity theory are most relevant.

\textbf{Table 2.} \textit{General relativistic calculation. Central
pressure, }$p\left( 0\right) ,$\textit{\ and central density, }$\rho \left(
0\right) ,$\textit{\ are in units }$7.2\times 10^{18}$\textit{\ kg m}$^{-3},$%
\textit{\ star radius, }$R,$\textit{\ in kilometers, mass, }$M,$\textit{\ in
solar masses and baryon number, }$N,$\textit{\ in units of solar baryon
number. \bigskip I report also the percent binding energy, }$BE$\textit{,
defined by the ratio }$100(N-M)/N$\textit{\ and the fractional surface red
shift, }$\Delta \lambda /\lambda =1/\sqrt{1-2M/R}-1$\textit{\ .}

$
\begin{array}{llllllll}
p\left( 0\right) & 0.01 & 0.04 & 0.1 & 1 & 10 & 100 & 1000 \\ 
\rho \left( 0\right) & 0.18 & 0.46 & 0.89 & 5.3 & 39 & 3.4\times 10^{2} & 
3.1\times 10^{3} \\ 
R & 11.9 & 9.6 & 8.3 & 5.8 & 5.2 & 6.6 & 6.6 \\ 
M & 0.67 & 0.72 & 0.70 & 0.55 & 0.39 & 0.40 & 0.44 \\ 
N & 0.69 & 0.74 & 0.73 & 0.55 & 0.36 & 0.37 & 0.42 \\ 
BE & 2.9\% & 3.4\% & 3.4\% & -0.73\% & -8.1\% & -8.0\% & -5.7\% \\ 
\Delta \lambda /\lambda & 0.094 & 0.13 & 0.15 & 0.18 & 0.13 & 0.11 & 0.12
\end{array}
$

Table 2 shows that both the mass, $M$, and the baryon number, $N$, increase
with increasing central density until $\rho \left( 0\right) \simeq 0.46,$
where $M\simeq 0.72$ $M_{\circ }$ (the $OV$ mass limit)$,$ and both $M$ and $%
N$ decrease after that. There are no equilibrium configurations, either
stable or unstable, with baryon number above $N\simeq 0.74.$ Actually for
every baryon number $N<0.74$ there are two equilibrium configurations, one
of them with $\rho \left( 0\right) <0.46$ and another one with $\rho \left(
0\right) >0.46,$ the latter having higher mass than the former. Furthermore,
as is shown in Table 2, stars with large central density have a negative
binding energy and therefore cannot be stable.

\end{document}